\documentclass{ws-rv9x6}
\usepackage{ws-rv-van}             
\usepackage{graphicx}
\usepackage{ws-index}             
\makeindex
\newindex{aindx}{adx}{and}{Author Index}       
\renewindex{default}{idx}{ind}{Subject Index}  

\newcommand{\vect}[1]{{\mathbf #1}}
\newcommand{\bra}[1]{\langle\left.{#1}\right|}
\newcommand{\ket}[1]{\left|{#1}\right.\rangle}
\newcommand{\eF}{\varepsilon_F}

\begin{document}

\chapter[The BCS--BEC Crossover]{The BCS--BEC Crossover\label{chapterParish}}

\author[M. M. Parish]{Meera M. Parish 
}

\address{
London Centre for Nanotechnology, University College London \\ 
Gordon Street, London WC1H 0AH, United Kingdom
}

\begin{abstract}
This chapter presents 
the crossover from the Bardeen--Cooper--Schrieffer (BCS) state of weakly-correlated pairs of fermions to the Bose--Einstein condensation (BEC) of diatomic molecules in the atomic Fermi gas. 
Our aim is to provide a pedagogical review of the BCS--BEC crossover,
with an emphasis on the basic concepts, particularly those that are not generally known or
are difficult to find in the literature. We shall not attempt to give an exhaustive survey of current research in the limited space here; where possible, we will direct the reader to more extensive reviews.
\end{abstract}

\body

\section{Introduction}\label{sec:intro_parish}
\begin{figure}
\centering
\includegraphics[width=0.8\linewidth]{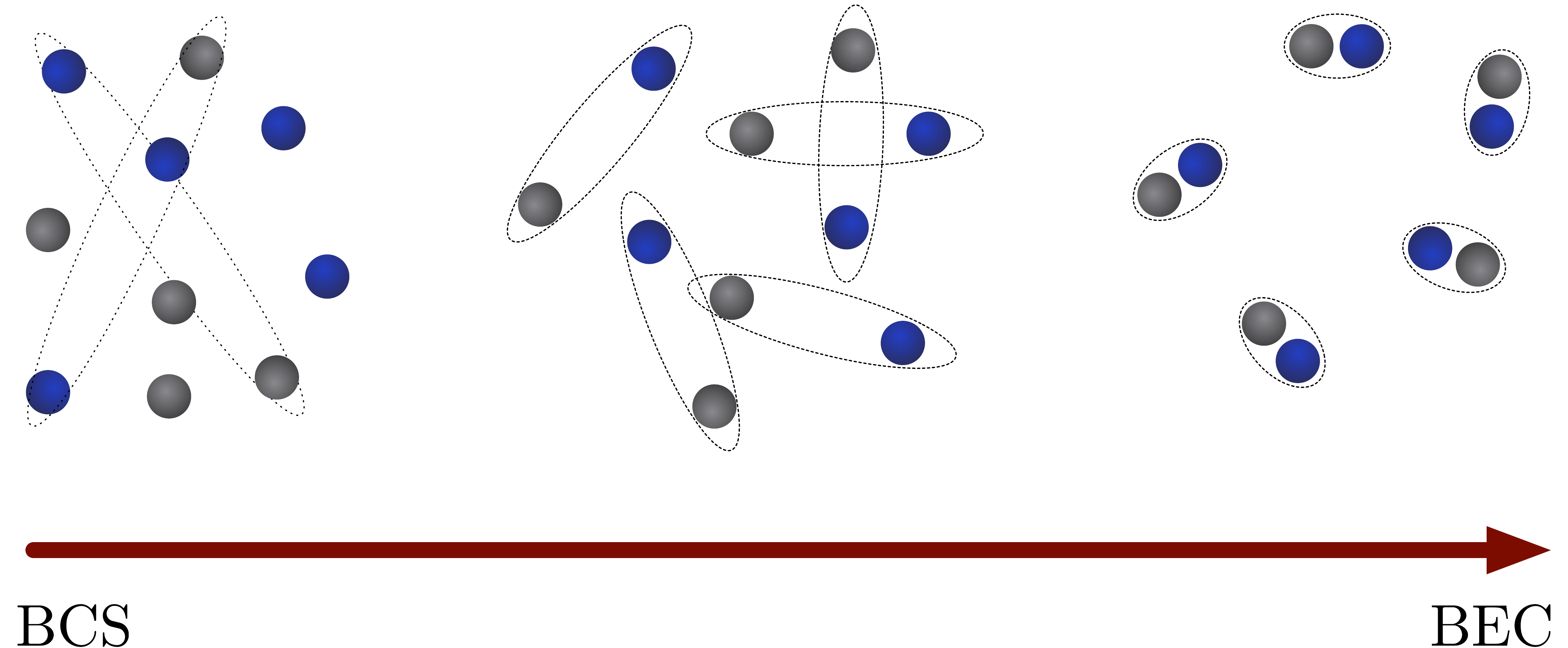}
\caption{Crossover from BCS to BEC regimes in a two-component Fermi gas.}
\label{fig0_parish}
\end{figure}

Ultracold atomic vapors provide a unique and tunable experimental
system in which to explore pairing phenomena, particularly in the
context of Fermi gases. 
A defining moment in the field was the successful realization of the
crossover from the BCS state of Cooper pairs to the Bose--Einstein
condensation (BEC) of diatomic
molecules~\cite{regal2004,zwierlein2004,chin2004,bourdel2004,kinast2004,zwierlein2005}.  
The purpose of this chapter is to review the basic concepts of this
BCS--BEC crossover in atomic Fermi gases. 
The idea of the BCS--BEC crossover, in fact, predates cold-atom
experiments by several decades~\cite{eagles1969,Leggett1980}. 
Indeed, it is a generic feature of attractively interacting Fermi
gases and can thus occur (at least in principle) in a variety of
systems ranging from superconductors and excitons in semiconductors,
to neutron stars and QCD. 
However, thus far it has only been
unequivocally observed in the dilute atomic gas.
In all cases, the crossover is achieved by varying the length scale of
the pairing correlations (i.e.\ the `size' of the fermion pairs) with
respect to the interparticle spacing, as depicted in Fig.~\ref{fig0_parish}.
This clearly yields two ways to drive the crossover: by fixing the
interactions and changing the particle density, or by tuning the
interactions at fixed density.  
The former ``density-driven'' crossover is typical of Coulomb systems
like excitons~\cite{comte1982} where the interactions cannot be easily
altered and there is always a two-body bound state, 
while the latter ``interaction-driven'' crossover is achieved in
atomic gases via the use of the Feshbach resonance (see
Chapter~4
).  
The fact that there is a crossover rather than a phase transition is non-trivial to prove 
theoretically, 
but can be argued heuristically on the grounds that both limits are captured by the same wave function, as discussed below. 
Note, however, that for pairing at non-zero angular momentum, e.g., $p$-wave
pairing~\cite{Leggett1980}, 
there is in fact a \textit{phase transition} between the BCS and BEC regimes
at zero temperature rather than a crossover. Thus, this chapter will
be confined to a discussion of isotropic $s$-wave pairing only.

\section{The two-component Fermi gas}
For low energy, $s$-wave interactions, such as those found in the
cold-atom system, Pauli exclusion forbids scattering between identical
fermions and thus we require at least two species of fermions to
produce pairing.  The different species can correspond to different
hyperfine states of the same atom, or single hyperfine states of
different atomic species such as $^6$Li and $^{40}$K. 
The physics of pairing in a Fermi gas 
is best elucidated by considering a uniform, two-component
($\uparrow$, $\downarrow$) Fermi gas in three dimensions (3D),
described by the Hamiltonian: 
\begin{align}\label{eq:ham_parish}
\hat{H} -\mu \hat{N} = & \sum_{\vect{k}\sigma} \left(\epsilon_{\vect{k}} -\mu \right)
 \hat{c}^\dag_{\vect{k}\sigma}\hat{c}_{\vect{k}\sigma}
 + \frac{U}{V}
\sum_{\vect{k},\vect{k'},\vect{q}}
 \hat{c}^\dag_{\vect{k}\uparrow} \hat{c}^\dag_{\vect{k'}\downarrow}
 \hat{c}_{\vect{k'}+\vect{q}\downarrow}\hat{c}_{\vect{k}-\vect{q}\uparrow} 
\end{align}
where the spin $\sigma=\{\uparrow,\downarrow \}$, the momentum dispersion
$\epsilon_{\vect{k}} = \frac{\hbar^2\vect{k}^2}{2m}$, $V$ is the
system volume, $\mu$ is the chemical potential, and $U<0$ is the strength of an
attractive contact interaction. 
We will focus on the simplest case of equal masses $m_\uparrow =
m_\downarrow \equiv m$, 
since the qualitative behavior of the BCS--BEC crossover is not
expected to change for unequal masses.\footnote{At least not for small
mass imbalance. 
For sufficiently unequal masses, 
one eventually expects clustering and crystallization to compete with
the condensation of pairs.} 
Note that we require the chemical potential and thus the density of
each spin component to be equal -- imbalancing the spin populations
will frustrate pairing and produce a more complicated phase diagram
with both first- and second-order phase
transitions.~\cite{parish2007,sheehy2007}  
The interparticle spacing in the Fermi gas can be parameterized by the
Fermi momentum $k_F \equiv (6\pi^2 n)^{1/3}$, where $n=N/2V$ is the 3D
density of each component. 
In the absence of interactions, the ground state wave function is
$\prod_{|\vect{k}|<k_F} \hat{c}^\dag_{\vect{k}\uparrow} \hat{c}^\dag_{\vect{k}\downarrow} \ket{0}$, 
corresponding to a filled sphere in momentum space with radius $k_F$
and chemical potential $\mu$ given by the Fermi energy $\varepsilon_F
= \hbar^2 k_F^2/2m$. 

As discussed in earlier chapters such as Chapter~4 
the short-range interactions in 
3D dilute atomic gases are characterized by the $s$-wave scattering
length $a_S$. This can be related to the bare interaction $U$
via~\cite{Fetter1971} 
\begin{align} \label{eq:lipp_parish}
\frac{m}{4\pi\hbar^2 a_S} = \frac{1}{U}
+ \sum_{\mathbf{k}}^{\Lambda} \frac{1}{2\epsilon_{\mathbf{k}}} 
\end{align}
where $\Lambda$ is an ultraviolet cut-off that is physically related
to the inverse of the range of the interaction potential. Here, it is
assumed that the gas is sufficiently dilute (i.e.\ the collisions in
the gas are sufficiently low in energy) that the behavior is
insensitive to the microscopic details of the potential. For
degenerate gases, this corresponds to the condition $\Lambda \gg
1/a_S$, $\Lambda \gg k_F$. Formally, one can take the limits $U \to 0$,
$\Lambda \to \infty$ while keeping the left hand side of
Eq.~\eqref{eq:lipp_parish} fixed and finite. Removing the bare
interaction from the problem implies that the ground state only
depends on a single dimensionless parameter $k_Fa_S$. This can then be
varied  
by tuning $a_S$ with a Feshbach resonance, as described in
Chapter~4, 
where $1/a_S = 0$ signals the
appearance of a two-body bound state.  
The BCS and BEC regimes then correspond, respectively, to the limits
$1/k_Fa_S \gg -1$ and $1/k_Fa_S \gg 1$, while the ``crossover region''
can be defined as $k_F|a_S| > 1$. 

Other types of 
interactions, e.g., the dipole--dipole interactions considered in
Chapter~13, 
can  also be described by a scattering length, but a full characterization
will generally require additional length scales depending on the
structure and range of the effective potential.
The simplest extension 
is the case of the \textit{narrow} Feshbach resonance, where the
closed-channel-molecule component of the resonance becomes
significantly occupied and must therefore be included explicitly in
the BCS--BEC crossover~\cite{holland2001,timmermans2001}. 
We will examine this case in Section~\ref{sec:narrow}.

\section{Ground state and phenomenology} 
Considerable insight into the BCS--BEC crossover can be gained from using a simplified wave function for the paired ground state, 
corresponding to a mean-field description of pairing~\cite{eagles1969,Leggett1980}.  
In particular, it demonstrates how the BCS and BEC regimes are
smoothly connected in the ground state, and it provides a
qualitatively 
accurate picture of the whole crossover  
even in the strongly interacting 
unitarity regime $k_F|a_S|\gg 1$.

It is first instructive to consider the ground-state wave function in
the BEC limit $1/k_F a_S \gg 1$. Here, the size of the two-body bound
state is $a_S$ (recall from Chapter~4 
that there is a two-body bound state  when $a_S >
0$ with binding energy $\varepsilon_B = \hbar^2/ma_S^2$) and thus it is
much smaller than the interparticle spacing $\sim 1/k_F$. 
In this case, we can approximate the dimers as point-like bosons $b$
and  
the ground-state wave function can be written as a coherent state of
these bosons: $\ket{\Psi} = \mathcal{N} e^{\lambda \hat{b}^\dag_0} |
0 \rangle$, where $\mathcal{N}$ is a normalization constant and
$\lambda = \bra{\Psi} \hat{b}_0 \ket{\Psi}$ is the condensate order
parameter, i.e., $|\lambda|^2/V$ corresponds to the condensate
density. Of course, this assumes that the Bose gas is very weakly
interacting so that essentially all the bosons reside in the
condensate, but this is reasonable since the effective boson--boson
interactions tend to zero as $1/k_F a_S \to \infty$. Note, further,
that $\ket{\Psi}$ does not conserve particle number and it thus
corresponds to a condensate with a well-defined phase, unlike the
number state $(\hat{b}^\dag_0)^N | 0 \rangle$. It can be argued that
the coherent state is energetically favored over the number state in
the presence of weak repulsive
interactions~\cite{nozieres1995}. However, in practice they both yield
equivalent results for thermodynamic quantities and thus we can use
whichever is most convenient.  

\begin{figure}
\centering
\includegraphics[width=0.7\linewidth]{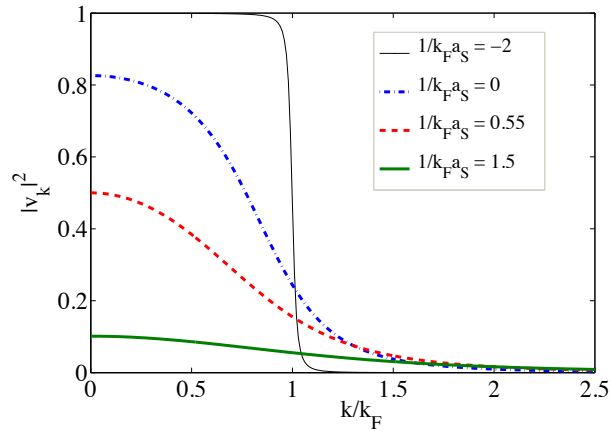}
\caption{Evolution of the momentum distribution $v_\vect{k}^2$ with interaction $1/k_Fa_S$ across the BCS--BEC crossover.}
\label{fig1_parish}
\end{figure}

Now, 
since each boson is composed of two fermions, we can write the boson
operator as $\hat{b}^\dag_\vect{q}
= \sum_\vect{k} \varphi_\vect{k} \hat c^\dag_{\mathbf{k}\uparrow} \hat
c_{\vect{q}-\mathbf{k}\downarrow}^\dag$, where $\varphi_\vect{k}$ is
the relative two-body wave function in momentum space. Inserting this
into $\ket{\Psi}$ then gives 
\begin{align} \label{eq:BCS_parish}
 |\Psi\rangle = \mathcal{N} e^{\lambda \sum_{\mathbf{k}} \varphi_\mathbf{k} \hat c^\dag_{\mathbf{k}\uparrow} \hat c_{-\mathbf{k}\downarrow}^\dag} | 0 \rangle
=
\prod_{\mathbf{k}} (u_\mathbf{k} + v_\mathbf{k}
 \hat c^\dag_{\mathbf{k}\uparrow} \hat c_{-\mathbf{k}\downarrow}^\dag) | 0 \rangle 
\end{align}
where $v_\vect{k}/u_\vect{k} = \lambda\varphi_\vect{k}$, $\mathcal{N}
= \prod_\vect{k} u_\vect{k}$, and we require $|u_\vect{k}|^2 +
|v_\vect{k}|^2 = 1$ for normalization. But Eq.~\eqref{eq:BCS_parish}
is nothing more than the celebrated BCS wave
function~\cite{schrieffer_book} which describes weakly bound pairs in
the limit $1/k_Fa_S \gg -1$. 
Thus, we see that the \textit{same} type of wave function describes
both the BCS and BEC limits.  
Indeed, 
we also recover the wave function for the non-interacting Fermi gas in
the limit $1/k_Fa_S \to -\infty$ 
by taking: 
\begin{align}\label{eq:step_parish}
  v_{\mathbf{k}} =
  \begin{cases}
   1, & \text{$|\mathbf{k}|< k_F$} \\
   0, & \text{$|\mathbf{k}|> k_F$} 
  \end{cases}
\end{align}
In the presence of a Fermi sea, arbitrarily weak attractive
interactions will generate pairing, 
in contrast to the two-body problem in a vacuum, 
which requires $a_S > 0$, i.e., a sufficiently strong attraction, for a bound pair to exist in 3D. 
At zero temperature, this leads to a condensate  
of strongly overlapping pairs, otherwise known as Cooper pairs, in the
BCS regime.  
Here, the sharpness of the Fermi surface is smeared out by the pairing
between fermions, but the majority of the fermions deep within the
Fermi sea remain unaffected, so that the momentum distribution
$\bra{\Psi}\hat{c}^\dag_{\vect{k}\uparrow}\hat{c}_{\vect{k}\uparrow} \ket{\Psi}=
|v_{\vect{k}}|^2$ still closely resembles a step function  (see
Fig.~\ref{fig1_parish}). Thus, the effect of exclusion is such that the pairing
correlations for Cooper pairs can be regarded as occurring in momentum
space rather than real space.  

To determine the ground state properties throughout the crossover, we
consider  
the free energy  
$\Omega=\bra{\Psi} \hat{H}
-\mu \hat{N} \ket{\Psi}$, which corresponds to the following in terms
of $v$, $u$: 
\begin{align}\label{eq:energy_parish}
\Omega = 2 \sum_\vect{k} (\epsilon_\vect{k} - \mu) |v_\vect{k}|^2 + \frac{U}{V} \sum_{\vect{k}\vect{k'}} v_\vect{k}^* u_\vect{k} v_\vect{k'} u_\vect{k'}^* + \frac{U}{V} \sum_{\vect{k}\vect{k'}} |v_\vect{k}|^2 |v_\vect{k'}|^2
\end{align}
Note that the factors $v$, $u$ only depend on the magnitude $k \equiv
|\vect{k}|$ since we are restricted to $s$-wave pairing. This also
means we can take $v$, $u$ to be real without loss of generality. 
The last term in Eq.~\eqref{eq:energy_parish} corresponds to the
lowest order mean-field Hartree term $U n^2$, which can be neglected
since it vanishes in the limit of short-range interactions, $U\to 0$. 

Minimizing\footnote{Hint: We must take the derivative of Eq.~\eqref{eq:energy_parish} with respect to, say,  $v_\vect{k}$ while keeping the constraint $u_\vect{k}^2 + v_\vect{k}^2 = 1$. The simplest way to do this is to define $v_\vect{k} = \sin\theta_\vect{k}$, $u_\vect{k} = \cos\theta_\vect{k}$, and then take $\partial\Omega/\partial\theta_\vect{k} = 0$. One should also check that $\partial^2\Omega/\partial\theta_\vect{k}^2 >0$ to verify that the stationary point correponds to a minimum.} 
$\Omega$ at fixed $\mu$ then yields the following condition for $v_\vect{k}$:
\begin{align}
2 (\epsilon_\vect{k} -\mu) u_\vect{k} v_\vect{k} + (u_\vect{k}^2 - v_\vect{k}^2) \frac{U}{V}\sum_{\vect{k'}} u_\vect{k'} v_\vect{k'} = 0 
\end{align}
In the limit 
$v_{\vect{k}} \to 0$, where the effects of Pauli exclusion should be
negligible, this reduces to the Schr\"odinger equation for the
two-body bound state with wave function $v_\vect{k}/\sqrt{N}$ and
binding energy $-2\mu$.  
Thus, in the BEC regime, $\mu \to -\varepsilon_B/2$ and
$v_{\vect{k}}$ becomes the two-body bound state wave function
$\varphi_\vect{k} \sim 1/(2\epsilon_\vect{k}+\varepsilon_B)$, as shown
in Fig.~\ref{fig1_parish}.  
More generally, we must solve the equations:
\begin{align}\label{eq:delta_parish}
\Delta & \equiv  \frac{U}{V} \sum_\vect{k} u_\vect{k} v_\vect{k} = -\frac{U}{V} \sum_\vect{k} \frac{\Delta}{2E_\vect{k}} \\ \label{eq:number_parish}
n & = \frac{1}{V} \sum_\vect{k} v_\vect{k}^2 = \frac{1}{2V} \sum_\vect{k} \left(1 -\frac{\epsilon_\vect{k}-\mu}{E_\vect{k}} \right) \\ \label{eq:dispersion_parish}
E_{\vect{k}} & = \sqrt{(\epsilon_\vect{k} - \mu)^2 + \Delta^2} 
\end{align}
where Eqs.~\eqref{eq:delta_parish} and \eqref{eq:number_parish} correspond, respectively, 
to the usual gap  and number equations.
We have also introduced the standard BCS order parameter $\Delta$,
which gives a measure of the pairing correlations in the condensate.  
In the BCS limit, it corresponds to the pair binding energy, as
discussed below, while in the BEC limit it reduces to a normalization
constant $\sim \varepsilon_F/\sqrt{k_Fa_S}$ for the two-body wave
function $v_{\vect{k}} \simeq \sqrt{N}\varphi_\vect{k}$. One can see this by noting that 
$v_{\vect{k}} \simeq \Delta/(2\epsilon_\vect{k}+\varepsilon_B)$ in the BEC limit and then using the density $n = \sum_\vect{k} v_{\vect{k}}^2/V$ to fix $\Delta$.  

\begin{figure}
\centering
\includegraphics[width=0.6\linewidth]{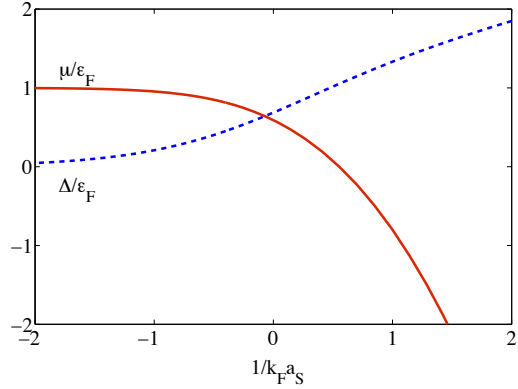}
\caption{Behavior of the chemical potential $\mu/\varepsilon_F$ and the order parameter $\Delta/\varepsilon_F$ as the interaction $1/k_Fa_S$ is varied.}
\label{fig2_parish}
\end{figure}

Figure~\ref{fig2_parish} depicts the evolution of
$\Delta/\varepsilon_F$ and $\mu/\varepsilon_F$ throughout the
crossover. We see that both quantities smoothly interpolate between
the BCS and BEC limits, as expected from the form of the wave
function \eqref{eq:BCS_parish}.  
Likewise, the momentum distribution $v_\vect{k}^2$ in
Fig.~\ref{fig1_parish} evolves continuously from a step-like function
to one  
spread out in momentum with increasing $1/k_Fa_S$. 
In the BCS regime $1/k_Fa_S \ll -1$, the chemical potential
$\mu \simeq \varepsilon_F$, while $\Delta$ tends to zero exponentially
as $1/k_Fa_S \to -\infty$,  
which is consistent with the existence of pairing for arbitrarily weak
interactions. Of course, the non-interacting state $\Delta = 0$ is
also a trivial solution of Eq.~\eqref{eq:delta_parish}, but one can
show that this always has a higher energy than the paired state, 
i.e., it corresponds to a \emph{maximum} rather than a minimum of $\Omega$.  
Note that we must vary $a_S$ to achieve the crossover if we want to
remain in the dilute limit $\Lambda \gg k_F$. For a density-driven
crossover where $a_S$ is fixed and $k_F$ is varied, we  
will always have $k_F a_S >0$. 
Thus, in order to access the BCS regime, we must eventually depart
from the universal curves in Fig.~\ref{fig2_parish} as
$k_F a_S \to \infty$ and instead have behavior that is sensitive to
the details of the interaction~\cite{parish2005_2}.

\subsection{Low energy excitations} \label{sec:exc}
The low-energy  
excitations of the ground state wave function~\eqref{eq:BCS_parish}
are best elucidated by considering an alternative derivation of the
mean-field
equations \eqref{eq:delta_parish}--\eqref{eq:dispersion_parish}. We can
equivalently define  
%
$\Delta = \frac{U}{V}\sum_{\vect{k}} \langle \hat{c}_{-\vect{k}\downarrow}\hat{c}_{\vect{k}\uparrow} \rangle$ 
%
and then take the fluctuations about this expectation value to be small, i.e.,
\begin{align}
 \sum_{\vect{k},\vect{q}} \hat{c}_{\vect{q}-\vect{k}\downarrow}\hat{c}_{\vect{k}\uparrow} = V\frac{\Delta}{U}\delta_{\vect{q}0} +\hat\eta_\vect{q}
\end{align}
where $\hat\eta_\vect{q}$ is small. 
Inserting this into Eq.~\eqref{eq:ham_parish} and expanding up to
first order in $\hat\eta_\vect{q}$ then yields the  
reduced mean-field Hamiltonian:
\begin{align}\label{eq:MF_ham_parish}
  \hat{H}_{MF}
  = - V \frac{\Delta^2}{U} + \sum_{\vect{k}}
   (\epsilon_\vect{k} -\mu)
  + \sum_{\vect{k}} \hat\psi_{\vect{k}}^\dag
  \begin{pmatrix}
    \epsilon_{\vect{k}} - \mu & \Delta\\
    \Delta & \mu -\epsilon_{\vect{k}}
  \end{pmatrix}
  \hat\psi_{\vect{k}}  ,
\end{align}
where $\hat\psi_{\vect{k}}^\dag = (\hat{c}_{\vect{k} \uparrow}^\dag ,
\hat{c}_{-\vect{k} \downarrow})$.
We now diagonalise the Hamiltonian $\hat{H}_{MF}$ using the standard Bogoliubov
transformation~\cite{Schuck_book}
\begin{align*}
  \begin{pmatrix} \hat\gamma_{\vect{k} \uparrow} \\
  \hat\gamma_{-\vect{k} \downarrow}^\dag \end{pmatrix} 
= 
  \begin{pmatrix}
    u_{\vect{k}} &- v_{\vect{k}}\\
    v_{\vect{k}} & u_{\vect{k}}
  \end{pmatrix}
  \begin{pmatrix} \hat{c}_{\vect{k} \uparrow} \\ \hat c_{-\vect{k}
      \downarrow}^\dag \end{pmatrix} 
\end{align*}
where  $u_\vect{k}$, $v_{\vect{k}}$ are the same as those defined previously.  
This yields the Hamiltonian $\hat{H}_{MF} =  \langle \hat{H}_{MF} \rangle + \sum_{\vect{k}\sigma} E_\vect{k} \hat\gamma^\dag_{\vect{k} \sigma} \hat\gamma_{\vect{k} \sigma}$ and ground state energy
%
$\Omega = \langle \hat{H}_{MF} \rangle = - V \frac{\Delta^2}{U} + \sum_{\vect{k}}
   (\epsilon_\vect{k} -\mu - E_\vect{k})$.
Thus,  $\hat\gamma^\dag_{\vect{k} \sigma}$ is the creation operator for (fermionic) quasiparticle excitations and $E_\vect{k}$ is the corresponding excitation energy. Since the ground state wave function 
$\ket{\Psi}$ is such that $\hat\gamma_{\vect{k} \sigma} \ket{\Psi} =
0$, we must have
$\ket{\Psi} \propto \prod_{\vect{k}\sigma} \hat\gamma_{\vect{k} \sigma} \ket{0}$
and this is indeed equivalent to Eq.~\eqref{eq:BCS_parish}, since we have $\prod_{\vect{k}} \hat\gamma_{\vect{k} \uparrow} \hat\gamma_{-\vect{k} \downarrow} \ket{0} = \prod_{\vect{k}} v_{\vect{k}} 
(u_\mathbf{k} + v_\mathbf{k}
 \hat c^\dag_{\mathbf{k}\uparrow} \hat c_{-\mathbf{k}\downarrow}^\dag) \ket{0}$.
Moreover, we recover the gap equation \eqref{eq:delta_parish} and number
equation \eqref{eq:number_parish} by taking
$\partial\Omega/\partial\Delta = 0$ and $N =
-\partial\Omega/\partial\mu$, respectively. Note that the solution once again corresponds to minimizing the grand potential $\Omega$: the gap equation gives the condition for a stationary point, so in principle one must also calculate $\partial^2\Omega/\partial\Delta^2$ to assess whether or not it is a minimum. In practice, one can often guess this from the number of stationary points when $\Omega$ is bounded from below, e.g., if there are only two stationary points, one at $\Delta=0$ and one at $\Delta_0$, then $\Delta_0$ corresponds to the global minimum.

The form of $E_\vect{k}$ in Eq.~\eqref{eq:dispersion_parish} shows
that there is always an energy gap in the quasiparticle spectrum, and
this can be identified with (half of) the pair binding energy -- the
factor of a half comes from the fact that a broken pair involves two
quasiparticles, e.g.,
$\hat\gamma^\dag_{\vect{k} \uparrow}\hat\gamma^\dag_{-\vect{k} \downarrow}\ket{\Psi}$. In
the BCS limit, the minimum energy occurs at $\epsilon_\vect{k}
=\mu \simeq \varepsilon_F$ so that the gap is simply $\Delta$. By
contrast, in the BEC limit, the minimum energy is $\sqrt{\mu^2
+ \Delta^2} \simeq \varepsilon_B/2$, i.e., the pair binding energy is
$\varepsilon_B$, as expected. 
A good discussion of the nature of these quasiparticles is contained in Ref.~\refcite{schrieffer_book}.

\begin{figure}
\centering
\includegraphics[width=0.85\linewidth]{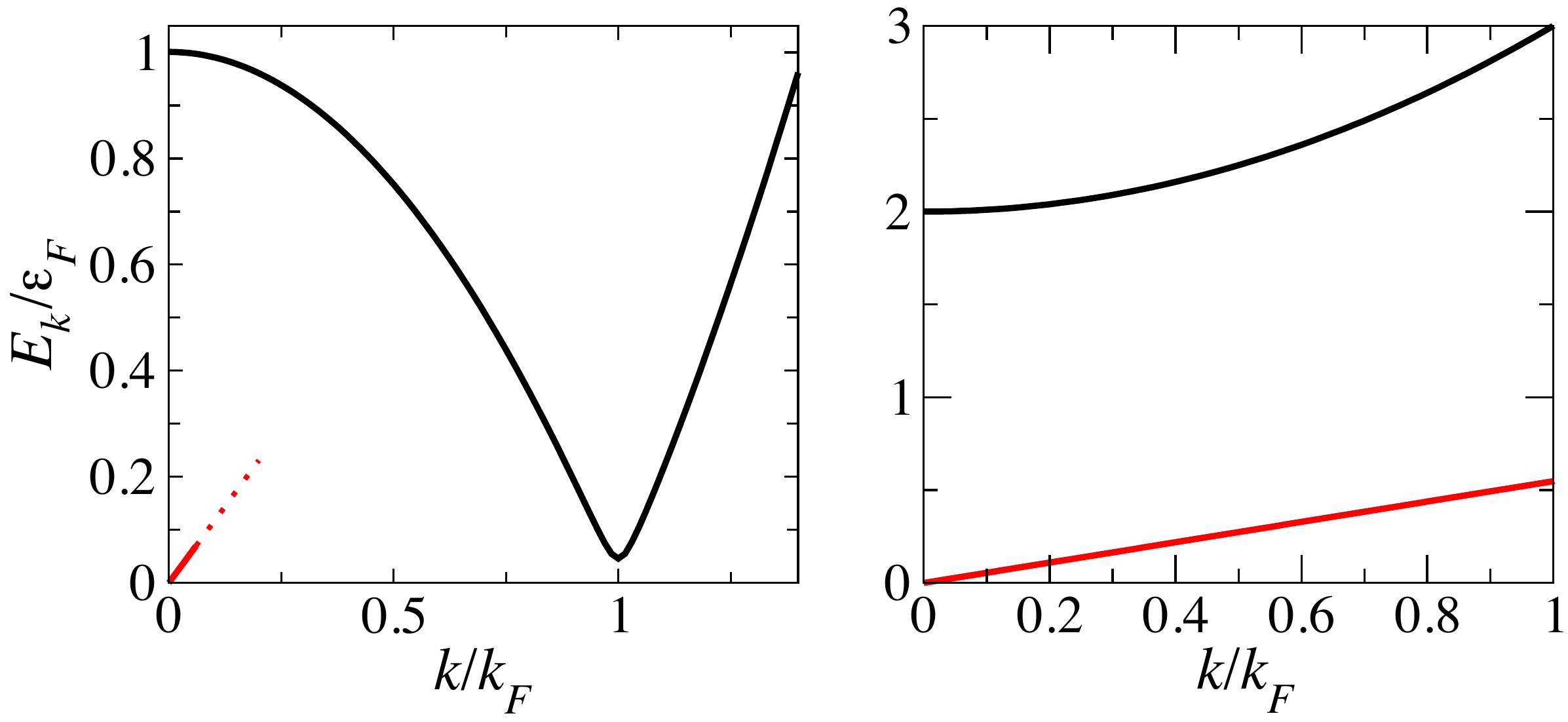}
\caption{Schematic depiction of the excitation dispersions in the BCS (left) and BEC (right) regimes. The black curves correspond to fermionic quasiparticle excitations, while the red lines are the linear Bogoliubov collective modes. Note that this latter excitation is damped for energies above the pair binding energy. In the BEC regime, the energy gap in the fermionic quasiparticle dispersion becomes $\varepsilon_B/2$, which is shown in the figure for $\varepsilon_B/\varepsilon_F \simeq 4$.}
\label{fig_disp_parish}
\end{figure}

In addition to these fermionic excitations, there is a low energy
bosonic collective mode (a gapless goldstone mode)
associated with the fluctuations
$\hat\eta_\vect{q}$ surrounding the mean-field order parameter
$\Delta$. It effectively involves the center-of-mass motion of the
pairs and its energy dispersion evolves into that of a free dimer in
the limit $1/k_Fa_S \to \infty$. The behavior of this excitation
throughout the crossover is perhaps best described within the
functional integral approach~\cite{altland2006}, where it corresponds
to Gaussian fluctuations around the mean-field saddle
point~\cite{engelbrecht1997}. In the BEC regime, where the pairing gap
is large, the bosonic collective mode becomes the only low-energy
excitation. The excitation energies in the BCS and BEC regimes are shown in Fig.~\ref{fig_disp_parish}.

\subsection{Crossover region and unitarity}
From the above analysis, we see that the system smoothly evolves from
the BCS regime, where there are primarily low-energy fermionic
excitations, to the BEC regime, where bosonic excitations dominate.  
However, in the crossover region $|k_Fa_S| > 1$, the pair size becomes of order
the interparticle spacing and thus the system 
can no longer 
be regarded as either a weakly interacting
Bose or Fermi gas. In particular, the unitarity limit $1/k_Fa_S = 0$
gives rise to a \textit{universal} strongly interacting Fermi
gas~\cite{ho2004} that is independent of any interaction length
scale. Therefore, at zero temperature, all thermodynamic quantities
only depend on the density via a universal constant $\xi$: for
instance, the chemical potential $\mu = \xi \varepsilon_F$ and the
total energy $E = \xi \frac{3}{5} \varepsilon_F N$. Ultracold 
gases have provided the first realization of such a unitary Fermi gas
and there has since been extensive work, both theoretical and
experimental, that we will not attempt to recapitulate here. We refer
the reader to Ref.~\refcite{Unitary_book2012} for an in-depth review 
of recent progress in the understanding of the unitary Fermi gas.

Another special point in the crossover region is that corresponding to $\mu =
0$. This marks a qualitative change in the fermionic quasiparticle
spectrum, since the minimum energy 
occurs at \textit{finite} momentum $k = \sqrt{2m\mu/\hbar^2}$ when $\mu >0$,
and  
at \textit{zero} momentum when
$\mu < 0$. Indeed, the point $\mu=0$ essentially signifies the
disappearance of a Fermi surface and it leads to a phase
transition for non-$s$-wave pairing~\cite{Leggett1980}. One may thus
define it as the crossover point between BCS- and BEC-type
behavior. As shown in Fig.~\ref{fig2_parish}, 
mean-field theory places it on the repulsive side of the Feshbach
resonance at $1/k_Fa_S \simeq 0.55$.

\subsection{Quantitative refinements}
While the mean-field approach has provided an intuitive and
qualitatively reasonable description of the BCS--BEC crossover, it is
not expected to be quantitatively accurate everywhere. Being
variational, it will at best provide an upper bound for the ground
state energy. The deficiencies of mean field theory are particularly apparent
at unitarity, where it neglects the strong many-body correlations
between pairs and significantly overestimates the energy: it predicts
$\xi \simeq 0.59$, whereas recent precision experiments on the unitary
Fermi gas~\cite{ku2012} yield $\xi \simeq 0.38$ in agreement with the
latest theoretical upper bound~\cite{forbes2011}.

Even in the weak-coupling BCS regime, the predicted mean-field
energy is incomplete since it neglects the interaction
energy of the normal Fermi liquid phase. Moreover, this interaction
energy dominates the 
correction to the ground state energy 
in the limit $1/k_Fa_S \to -\infty$ since it goes like $k_Fa_S$ to lowest
order, whereas the condensation energy $\sim \Delta^2$ is exponentially
small. 
There is also the so-called Gorkov--Melik-Barkhudarov
correction to the BCS order parameter $\Delta$ that arises from
the effects of induced interactions between fermions 
-- see, e.g., Ref.~\refcite{heiselberg2000}. 
This suppresses $\Delta$ by a constant factor, but the overall exponential dependence on $1/k_Fa_S$ is unchanged. 
 
In the BEC regime, 
we expect a weakly repulsive Bose gas that is characterized by an
effective dimer--dimer scattering length $a_{dd}$,  proportional to $a_S$. 
The energy shift due to this repulsion should give the
leading order correction to the chemical potential, i.e., $2\mu \simeq
-\varepsilon_B + \frac{2\pi\hbar^2 a_{dd}}{m} n$ for $1/k_Fa_S \gg 1$. 
The mean-field equations correctly recover this form for the repulsion
but with an incorrect scattering length, $a_{dd} = 2a_S$, which is an
overestimate compared with the 
exact result $a_{dd} \simeq 0.6 a_S$ obtained from four-body dimer--dimer
calculations~\cite{petrov2004}. To capture this result, one requires
a many-body wave function that incorporates four-body correlations exactly.

\section{Finite temperature}

\begin{figure}
\centering
\includegraphics[width=0.7\linewidth]{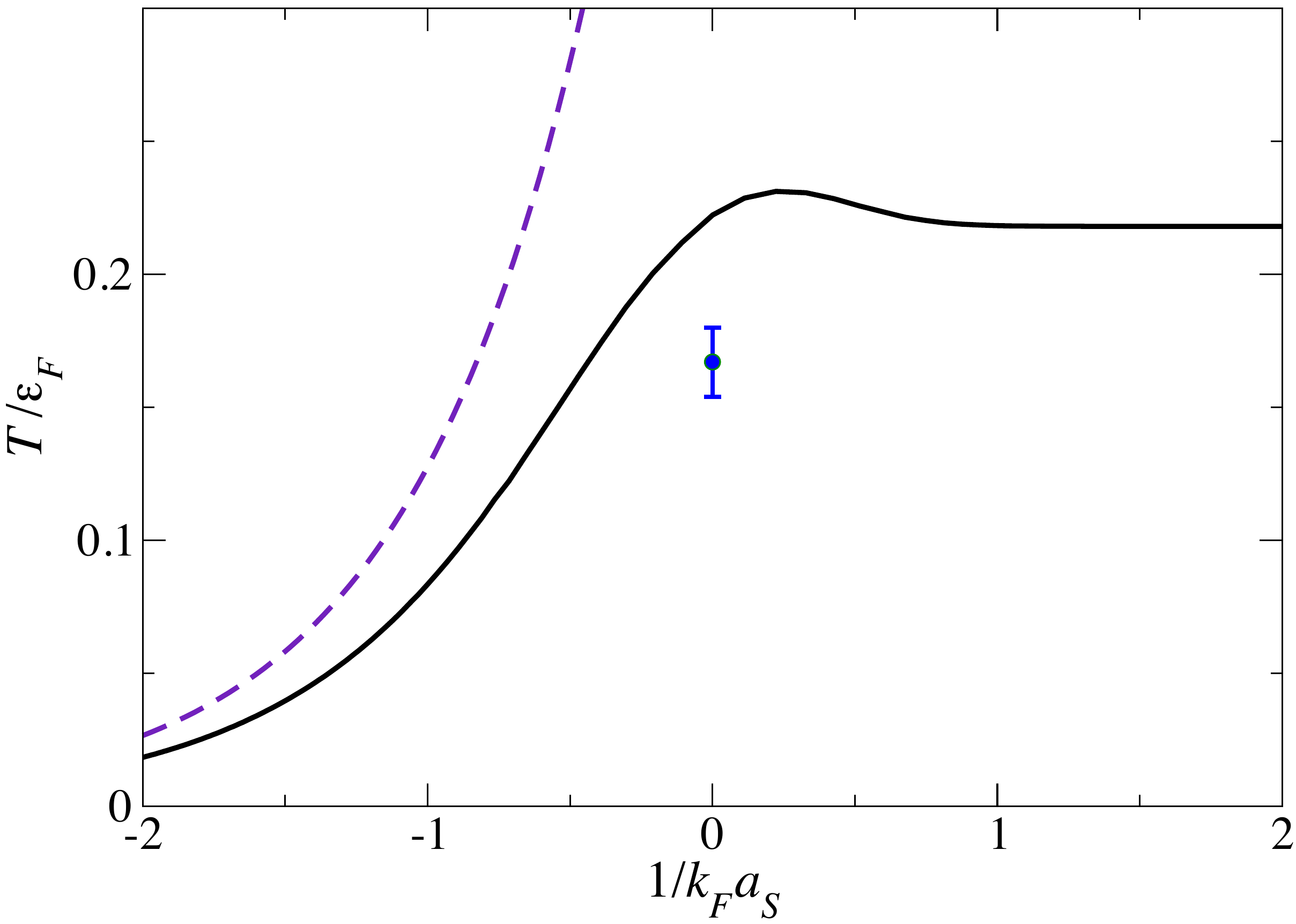}
\caption{Transition temperature $T_c$ for condensation throughout the BCS--BEC crossover, calculated using the Nozi\`eres--Schmitt-Rink approach~\cite{nozieres1985}. The dashed line marks the temperature $T^*$ around which pairs start to form. The filled circle marks the latest experimentally measured $T_c$
at unitarity~\cite{ku2012}, which is consistent with quantum Monte Carlo predictions~\cite{burovski2008,goulko2010}. 
}
\label{fig3_parish}
\end{figure}

We now turn to the effects of finite temperature $T$ on the 
BCS--BEC crossover. 
Here, the condensate of pairs will eventually be destroyed
for sufficiently large thermal fluctuations and, thus, the system
undergoes a continuous transition to a normal Fermi (Bose) gas in the BCS (BEC)
limit. Moreover, the transition temperature is determined by
the low-energy excitations of the condensate. 
Within the BCS regime
$1/k_Fa_S \ll -1$, where the pairing gap is small, pair condensation
essentially coincides with pair formation and therefore pair breaking
excitations will govern the transition. In this case, we can use 
the mean-field free energy\footnote{Note that this is the thermodynamic potential corresponding to the grand canonical ensemble, and it is often referred to as the \emph{grand potential}, which is distinct from other free energies such as Helmholtz, Gibbs, etc.} $\Omega(T)$ at finite temperature which is readily
obtained from Eq.~\eqref{eq:MF_ham_parish}:
\begin{align} \label{eq:grand_parish}
\Omega(T) = - V \frac{\Delta^2}{U} + \sum_{\vect{k}}
   (\epsilon_\vect{k} -\mu - E_\vect{k})
   - 2 k_B T \sum_\vect{k} \ln \left( 1 + e^{-E_\vect{k}/k_B T} \right)
\end{align}
where the BCS order parameter $\Delta(T)$ is now a function of temperature. 
With increasing temperature, $\Delta(T)$ becomes smaller and smaller, so that we can eventually expand Eq.~\eqref{eq:grand_parish} as follows: $\Omega(T) = \alpha \Delta^2 + \gamma \Delta^4 + ...$
The transition temperature $T_c$ then satisfies the condition
$\alpha \equiv \left. \frac{\partial \Omega }{\partial \Delta^2} \right|_{\Delta =
0} = 0$, i.e., it corresponds to the point where we no longer have a minimum at $\Delta \neq 0$. 
This yields
$T_c \sim \Delta(0)$ (we set $k_B=1$) and thus $T_c/\varepsilon_F$ goes to zero
exponentially when $1/k_Fa_S \to -\infty$, as shown in Fig.~\ref{fig3_parish}. 
Moving away from the BCS limit, the destruction of the condensate occurs before the loss of pairing and thus  
the critical temperature for pairing $T^*$ given by mean-field theory no longer coincides with $T_c$. Towards the BEC regime, $T_c$ is primarily determined by the bosonic collective modes, and in the limit $1/k_Fa_S
 \to \infty$, $T_c/\varepsilon_F$ saturates to the transition temperature for a non-interacting BEC, where $T_c/\varepsilon_F \simeq 0.218$.
In practice, it is difficult to model the evolution of $T_c$ between the BCS and BEC limits in a controlled fashion. The Nozi\`eres--Schmitt-Rink approach~\cite{nozieres1985} of including Gaussian fluctuations around the mean-field saddle point provides the simplest way of interpolating between the two limits~\cite{sademelo1993}. 
Even though it overestimates $T_c$ around unitarity compared to quantum Monte Carlo predictions~\cite{burovski2008,goulko2010}, 
it does correctly capture many of the qualitative features, such as the increase in $T_c/\varepsilon_F$ as we move away from the BEC limit and the maximum just before unitarity (see Fig.~\ref{fig3_parish}). For a survey of other theoretical methods, see  
Refs.~\refcite{bloch2008} and \refcite{giorgini2008}.

\section{Experiment}

\begin{figure}
\centering
\includegraphics[width=0.85\linewidth]{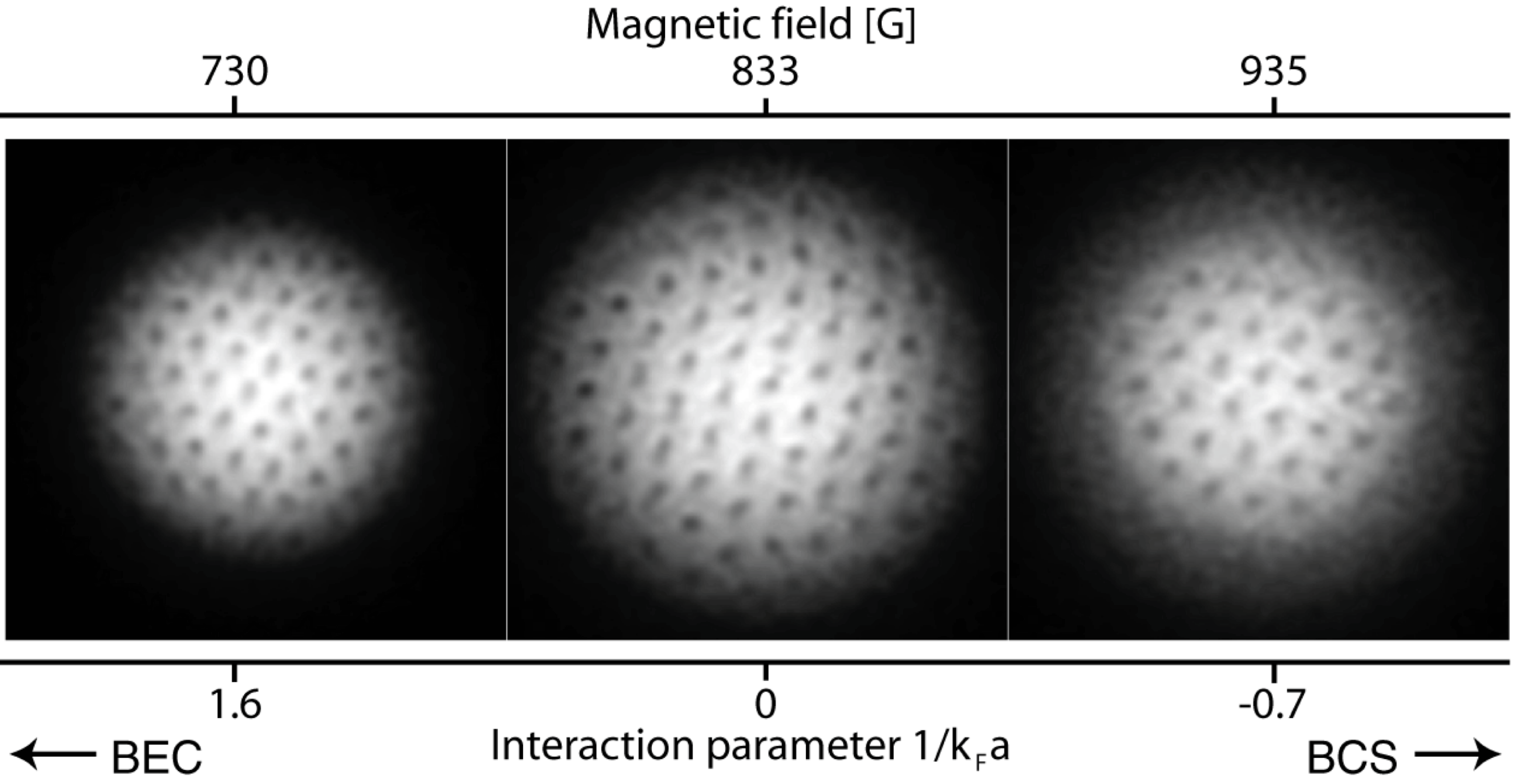}
\caption{Observation of vortices across the BCS--BEC crossover in  
Ref.~\refcite{zwierlein2005}.
}
\label{fig4_parish}
\end{figure}

The creation of ultracold atomic gases has meant that the Hamiltonian~\eqref{eq:ham_parish} can be realised directly in experiment and is more than just a useful toy model. Moreover, it is possible to access low enough temperatures that one can effectively extrapolate to the zero temperature behavior -- for instance, one can access the universal constant $\xi$. 
The condensate fraction throughout the crossover can be probed using time of flight measurements by first transferring all the pairs into molecules -- see Ref.~\refcite{regal2004}.
The pairing gap can also be measured~\cite{chin2004} using the RF spectroscopy described in Chapters~10 
and 11. 
More recently, momentum-resolved spectroscopy has allowed the quasiparticle excitation spectrum to be imaged directly, and evidence of pairing above $T_c$ has been observed~\cite{gaebler2010}.
There have also been increasingly better measurements of $T_c$ in the crossover region. The latest estimate~\cite{ku2012} at $1/k_Fa = 0$ is shown in Fig.~\ref{fig3_parish}.
Finally, there are the observations of quantized vortices~\cite{zwierlein2005}, as shown in Fig.~\ref{fig4_parish}, and second sound~\cite{sidorenkov2013}, both hallmarks of superfluidity. 
The dynamics of the Fermi gas and superfluidity are topics we have not touched upon here in this chapter, but 
a good introduction can be found in Ref.~\refcite{pethick_book}, in addition to  
the reviews of Refs.~\refcite{Unitary_book2012}, \refcite{bloch2008}, and \refcite{giorgini2008}. 

\section{Narrow Feshbach resonances} \label{sec:narrow}
In reality, the Feshbach resonance used to tune the interatomic interactions involves a closed channel component, as explained in Chapter~4. 
The minimal model to capture this is the two-channel Hamiltonian,  
which is obtained by replacing the ``single-channel'' interaction term $\frac{U}{V} \sum_{\vect{k},\vect{k'},\vect{q}} \hat{c}^\dag_{\vect{k}\uparrow} \hat{c}^\dag_{\vect{k'}\downarrow}
 \hat{c}_{\vect{k'}+\vect{q}\downarrow}\hat{c}_{\vect{k}-\vect{q}\uparrow}$ in Eq.~\eqref{eq:ham_parish} with:
\begin{align}
 \sum_\vect{k} \left(\frac{\epsilon_\vect{k}}{2}+\nu_0 -2\mu\right) \hat{d}^\dag_\vect{k} \hat{d}_{\vect{k}} + \frac{g}{\sqrt{V}} \sum_{\vect{k},\vect{q}} (\hat{c}^\dag_{\vect{k}\uparrow} \hat{c}^\dag_{\vect{q}-\vect{k}\downarrow} \hat{d}_\vect{q} + h.c.)
\end{align}
where $\hat{d}$ denotes a closed channel bosonic molecule with mass $2m$, $\nu_0$ is the ``bare'' detuning of the closed channel, and $g$ is the coupling between channels. Including the closed-channel boson explicitly leads to the modified mean-field ground state wave function $e^{\sqrt{N_d}\hat{d}^\dag_0}\ket{\Psi}$, where $N_d$ gives the number of closed channel molecules. Performing the mean-field analysis again, one can show that this effectively amounts to replacing $U$ with $g^2/(2\mu-\nu_0)$ in the equation for $\langle \hat{H}_{MF} \rangle$, with $N_d/V = \Delta^2/g^2$.
Likewise, we can relate these quantities to the scattering length $a_S$ by taking the zero energy limit $\mu=0$ and inserting $-g^2/\nu_0$ in place of $U$ in Eq.~\eqref{eq:lipp_parish}. This yields $-g^2/\nu = 4\pi\hbar^2a_S/m$, where $\nu \equiv \nu_0 -g^2 \sum_\vect{k}^\Lambda \frac{1}{2\epsilon_\vect{k}}$ is the renormalized physical detuning. The single-channel model is formally recovered by sending the closed channel off to infinity, i.e., by taking the limits $g\to\infty$, $\nu \to \infty$ while keeping $-g^2/\nu$ fixed.   
When $g$ is finite, there is an additional (inverse) length scale $l^{-1}\sim m^2 g^2/\hbar^4$ which defines the width of the resonance: for a broad resonance, we have $1/k_F l \gg1$, and the BCS--BEC crossover is well described by a single-channel model, while $1/k_F l \ll 1$ corresponds to a narrow resonance. In principle, the BCS--BEC crossover in cold atoms involves a superposition of both open-channel fermions and closed-channel bosons, with an increasing closed-channel component as we move towards the BEC side ($\nu<0$). 
However, experiments on the crossover have thus far only involved broad Feshbach resonances where the closed channel fraction is negligible~\cite{partridge2005}. 
Indeed, this 
has allowed experiments to access the unitary Fermi gas, since 
a significant $k_F l$ would have introduced  
an extra interaction length scale.

From a theory point of view, the narrow Feshbach resonance is more tractable because it provides a small parameter $1/k_F l$ throughout the crossover. Indeed, it can be shown that mean-field theory becomes a controlled approximation when $1/k_Fl \ll 1$, since corrections to the mean-field result are essentially perturbative in $1/k_F l$~\cite{gurarie2007}. Thus, in this case, the mean field approximation for the ground state and the Nozi\`eres--Schmitt-Rink approach to $T_c$ are \textit{quantitatively} accurate. 
To conclude this section, we note that even for a broad Feshbach resonance, the diluteness of the gas and the fact that the interparticle spacing is much larger than the range of the interactions can be used to constrain some properties of the many-body system, such as the pair correlations at short distances, leading to the concept of the so-called Tan ``contact'' \cite{tan2008,zhang2009}.

\section{Attractive Fermi Hubbard model}\label{sec:attr_hubb}
The BCS--BEC crossover may also be extended to the situation where there is 
an optical lattice. For a 3D square lattice, we can describe it by a Fermi Hubbard model, as discussed in Chapter~3: 
\begin{align}\label{eq:lattice_parish}
\hat{H}_{latt} = 
-J \sum_{\langle i,j \rangle,\sigma} \left( \hat{c}_{i,\sigma}^\dag \hat{c}_{j,\sigma} + h.c.\right) + U \sum_{i} \hat n_{i\uparrow} \hat n_{i\downarrow}
\end{align}
where $\langle i,j \rangle$ specifies nearest neighbor hopping between sites in the lattice, 
$J$ is the hopping energy and $U$ now corresponds to an attractive onsite interaction. 
Transforming to momentum space, one can derive a mean-field energy that resembles 
Eq.~\eqref{eq:energy_parish}, but with $\epsilon_\vect{k}$ replaced by the tight-binding dispersion 
$2J (3 - \cos (k_x a) - \cos (k_y a) - \cos (k_z a) )$, and momenta restricted to the first Brillouin zone $|k_x|, |k_y|, |k_z| \leq \pi/a$, where $a$ is the lattice spacing. Note, further, that $U$ is finite in the lattice case\footnote{Note that the high momentum cut-off in the lattice is set by $\pi/a$, so we do not have the limit $\Lambda \to \infty$, $U\to 0$ as in Eq.~\eqref{eq:lipp_parish}.} and so the Hartree term in Eq.~\eqref{eq:energy_parish} cannot be formally neglected -- in practice, it leads to a constant shift $U n$ of the chemical potential, and it means that the interaction energy of the normal Fermi liquid phase is included in the BCS mean-field theory, unlike in the continuum case without the lattice.

The extra length scale provided by the lattice means that the crossover now depends separately on the density, defined by the dimensionless parameter $\varepsilon_F/J$,~\footnote{In the lattice, we define $\varepsilon_F$ to be the chemical potential of the non-interacting Fermi gas with the same density $n$.} and the dimensionless interaction $|U|/J$. Moreover, there is a maximum density of $n=1$ particle per site for each spin, corresponding to $\varepsilon_F = 12J$. In this case, the system is simply a band insulator. 
For low densities $\eF \ll 12J$, the system behaves similarly to the continuum case in the BCS limit where the  interactions are weak, $|U|/J < 1$. By increasing the interactions, we eventually obtain a two-body bound state at $|U|/J \simeq 7.9$. The two-body binding energy $\varepsilon_B$ is given by the equation:
\begin{align}
\frac{1}{U} = \sum_\vect{k} \frac{1}{4J (3 - \cos (k_x a) - \cos (k_y a) - \cos (k_z a) ) +\varepsilon_B} 
\end{align}
However, once $|U|/J \gg 12J$, the size of the bound state is of order the lattice spacing $a$, with $\varepsilon_B \simeq U$, and the effects of the lattice become apparent.
In this regime, the size of the dimer is essentially constant (it cannot be smaller than $a$) and the effect of increasing $|U|$ is to localize the dimer in the lattice. To see this, one can perform second-order perturbation theory on Eq.~\eqref{eq:lattice_parish} for small $J/|U|$ to find that the hopping energy of a dimer is approximately $J^2/|U|$. Thus, the hopping goes to zero as $|U| \to \infty$.

This feature will strongly impact the BEC regime of the Hubbard model. While we still expect the system to tend towards a non-interacting BEC at zero temperature, the critical temperature $T_c$ scales with the dimer hopping energy, i.e., $T_c \sim J^2/|U|$, and it will thus approach zero instead of saturating like in Fig.~\ref{fig3_parish}, owing to the localization of bosonic dimers in the lattice.
Thus, $T_c$ tends to zero in both the BCS and BEC limits, with a pronounced maximum in between.
A discussion of the lattice case is also contained in Ref.~\refcite{Randeria1995}.

Another peculiarity of the Hubbard model is that it possesses particle-hole symmetry at half-filling, $\eF = 6J$. Thus, the regime $\eF > 6J$ corresponds to a BCS--BEC crossover of holes rather than particles, and the hole system at $\eF = 6J +\delta$ is equivalent to the particle system at $\eF = 6J -\delta$ (ignoring the Hartree term in the chemical potential).
However, a limitation of the Hubbard model is that it neglects the higher bands in the optical lattice, 
which become important when one approaches the Feshbach resonance and the interactions are strong. 
In particular, at unitarity $1/a_S =0$, the interactions scale with the lattice depth and thus can never be made small with respect to the band gap. Moreover, once $a>a_S>0$, the inclusion of higher bands yields 
dimers that are smaller than the lattice spacing. This makes it challenging to describe experiments on fermions in an optical lattice in the unitary regime~\cite{chin2006}.

\section{Concluding remarks}
While the underlying idea of the BCS--BEC crossover is quite simple to state, there are a surprising variety of subtleties that lead to the rich many-body physics outlined in this chapter. The elegant simplicity of the crossover also hides the fact that it is not \textit{a priori} obvious that such a system is even \emph{stable}. 
For the strong attractive interactions considered here, one runs the risk of triggering a collapse of the system into another phase, e.g., crystallization, rather than generating strong pairing. Indeed, this is 
a hidden conundrum that plagues many theories of high temperature superconductivity.
The fact that the cold atomic system can produce a (metastable) Fermi superfluid with the highest known $T_c$ compared to $\eF$ is because the inelastic decay processes leading to the loss of the gas are slower than the elastic collisions that are required for thermalization~\cite{petrov2004}. This can be even more pronounced in optical lattices, such as the 3D square lattice discussed in Section.~\ref{sec:attr_hubb}, or  low-dimensional geometries, which are currently an active area of research in cold atoms.
It remains to be seen whether the BCS--BEC crossover can be engineered in other condensed matter systems.

\section*{Acknowledgments}
I am grateful to Francesca Marchetti and Jesper Levinsen for fruitful discussions, and to 
Martin Zwierlein for providing me with the experimental figure.
This work was supported by the EPSRC under Grant No.\ EP/H00369X/2.

\bibliographystyle{ws-rv-van}
\bibliography{ChapterParish}

\printindex                         

\end{document}